\documentclass[aps, prb, twocolumn, groupedaddress,floatfix]{revtex4-1}
\usepackage{graphicx}
\usepackage{amsmath}
\usepackage{color}

\begin{document}

\title{Impurities and Landau level mixing in a fractional quantum Hall state in a flatband lattice model}
\author{Topi Siro}
\author{Mikko Ervasti}
\author{Ari Harju}

\affiliation{COMP Centre of Excellence, Department of Applied Physics,
  Aalto University, Helsinki, Finland}

\date{\today}

\begin{abstract}
We study the toplogical checkerboard lattice model around the
$\nu=\frac{1}{3}$ fractional quantum Hall phase using numerical exact
diagonalization without Landau level projections. We add local
perturbations, modified hoppings and on-site potentials, and observe
phase transitions from the fractional quantum Hall phase to metallic
and insulating phases when the strength and number of impurities is
increased. In addition to evaluating the energy spectrum, we identify
the phase diagrams by computing the topological Chern number of the
many-body ground state manifold, and we show how the ground states
lose their correlations due to the impurities by evaluating the
spectrum of the one-body reduced density matrix. Our results show that
the phase transition from the fractional quantum Hall phase to the
metallic phase occurs for both impurity hoppings and
potentials. Strong impurity hoppings cause a further transition into
the insulating state, regardless of the sign of the hopping, when
their density is high enough. In contrast, the same happens only for
attractive potentials.  Furthermore, the mixing to the higher band in
a two-band model, generally denoted as Landau level mixing, is
measured concluding that the lowest Landau level projection works well
even with remarkably strong interactions and in the presence of
impurities.
\end{abstract}


\maketitle

\section{Introduction}

There has been great interest recently in topological lattice models
that support states analogous to the ones in the integer and the
fractional quantum Hall (FQH) effects \cite{Ezawa_2000,
  Wen_2004}. Since the Haldane honeycomb\cite{Haldane_1988} model,
various topological lattice models have been introduced, most notably
ones without a net external magnetic field and with flat energy bands
arising from realistic hopping parameters \cite{Tang_2011,
  Neupert_2011, Sun_2011, Wu_2012, Bernevig_2012,
  Regnault_2011,Scaffidi_2012}. In these systems the time-reversal
invariance is broken to achieve quantized non-zero Hall conductivity,
characterized by the non-zero band Chern number
\cite{Thouless_1982}. It has been shown that filling such topological
bands according to the standard $\nu=\frac{1}{3}$ and
$\nu=\frac{1}{5}$ filling fractions, certain interactions stabilize
the FQH phase \cite{Sheng_2011}.

The FQH states are robust against weak local perturbations due to the
large energy gap between the ground states and the excited states, and
due to the topological order present in the ground states induced by
the interactions. Therefore, weak disorder should not break the FQH
phase. However, when increasing the number and strength of the local
impurities, at some point the system becomes a normal metal or an
insulator\cite{Sheng_2003}. This has been studied in a topological
lattice model with impurity on-site potentials positioned in a single
line, showing quantum phase transitions from FQH phase to a metallic
phase and with even stronger impurities to an insulating phase
\cite{Yang_2012}. However, it is not yet well understood how the
positioning of the impurities, and the types of impurities affect the
phase diagram. The impurities also change the effective filling
fraction of the rest of the system, which could have a pronounced
effect to break the FQH phase in finite systems.

The FQH topological phases can be characterized by the topological
ground state degeneracy \cite{Wen_1990}, and furthermore by the
quasiparticle properties \cite{Arovas_1984, Wen_1991, Wen_2004}. In
principle, impurities attract or repel the quasiparticles. One can
even hope to braid quasiparticles by moving the impurities, which has
been explicitly done in a numerical simulations of bosonic models
\cite{Kapit_2012,Diaz_2012}. Therefore, it is also interesting to
study the effect of individual impurites, especially by observing the
changes in the ground states as a function of the impurity strength.

In this work, we study the effect of impurities on the two-band
topological checkerboard lattice model around the $\nu=\frac{1}{3}$
FQH phase \cite{Sheng_2011}. We consider a variable number of
impurities, local perturbations on the nearest-neighbor hoppings, and
on-site potentials, with increasing and decreasing values. The
impurities are distributed as far away from each other as
possible. This setup makes it possible to study two kinds of disorder
effects, by keeping the number, or alternatively, the strengths of the
impurities fixed, while modifiying the other. The quantum phase
diagrams are deduced by evaluating the energy spectrum, topological
degeneracy, the many-body ground state manifold Chern number, and the
spectrum of the one-body reduced density matrix (1-RDM).  

The systems
are solved using numerical exact diagonalization method in the full
state space. This allows us to measure how much the two
non-interacting bands mix in the interacting system, generally denoted
as Landau level mixing. Due to memory constraints of the exact diagonalization method, we are only able to study quite small finite systems. 
However, as shown by previous studies\cite{Sheng_2011,Yang_2012}, interesting topological features can be observed 
despite the finite size of the lattice. The results for these small systems are also interesting from the point of view of the potential 
experimental realizations of flatband lattice models with ultracold atoms in optical lattices. 

\section{Model and Methods}

\begin{figure}[thb]
 \includegraphics[width=0.35\textwidth]{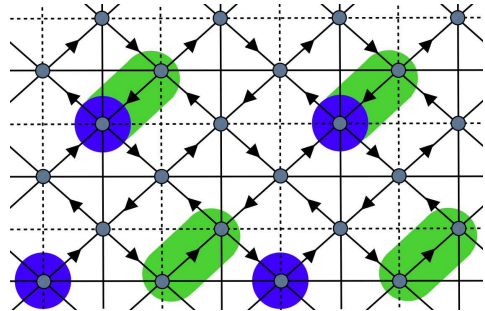}
  \caption{\label{fig1} (Color online) The checkerboard lattice with $4\times3$ unit cells that is used in the computations. 
  The arrows indicate the direction where the sign of the
    complex phase $\phi_{jk}$ in the nearest-neighbor hopping is
    positive. In the next-nearest-neighbor hopping, dashed and solid
    lines have opposite signs. In addition, there is a
    next-next-nearest-neighbor hopping not drawn in the figure for
    clarity. The green ovals and the blue circles indicate the
    positions of the impurity hoppings and impurity potentials,
    respectively. }
 \end{figure}

The two-band model on a checkerboard lattice was introduced in
Ref. \onlinecite{Sun_2011}.  The Hamiltonian reads
\begin{equation}
\label{Ham}
\begin{split}
H_0 = \hspace{6pt} &
\sum_{\left<j,k\right>}t_{jk}e^{i\phi_{jk}}c_j^{\dagger}c_k +
\sum_{\left<\left<j,k\right>\right>}t^{\prime}_{jk}c_j^{\dagger}c_k
\\ &+
\sum_{\left<\left<\left<j,k\right>\right>\right>}t^{\prime\prime}_{jk}c_j^{\dagger}c_k
+ \textnormal{H.c.},
\end{split}
\end{equation}
where $c_j^{\dagger}$ and $c_j$ are the fermionic creation and
annihilation operators for site $j$. The sums run over the
nearest-neighbor, next-nearest-neighbor and next-next-nearest-neighbor
sites, respectively, and the corresponding hopping amplitudes are $t$,
$t^\prime$, and $t^{\prime\prime}$.  The sign of the phase factor
$\phi_{jk}$ alternates between sites as seen in Figure \ref{fig1}. The
lowest band is almost flat and has a unit Chern number. To minimize
the dispersion in the lowest band, we use the values given in
Ref. \onlinecite{Sun_2011}: $\phi_{jk}=\pi/4$, $t=1$,
$t^\prime=\pm1/(2+\sqrt{2})$ and
$t^{\prime\prime}=1/(2+2\sqrt{2})$. Note that we have reversed the
sign of the Hamiltonian in Ref. \onlinecite{Sun_2011} to make the flat
band lowest in energy.

To obtain the $\nu=\frac{1}{3}$ FQH state, we add a nearest-neighbor
repulsive interaction,
\begin{equation}
\label{Hint}
H = H_0 + V\sum_{\left<j,k\right>}n_jn_k,
\end{equation}
where $n_j = c_j^{\dagger}c_j$ counts the number of electrons at site
$j$.  In Ref. \onlinecite{Sheng_2011} it was shown that for suitable
values of $V$, there is a threefold quasi-degenerate ground state
manifold (GSM), separated from the higher states by a large gap. We
introduce twisted boundary conditions,
\begin{equation}
\psi(x_j+L_j) = e^{i\theta_j}\psi(x_j),
\end{equation}
where $j$ indexes the space dimensions and $L_j$ is the length of the
system along direction $j$. The gap above the GSM remains open for all
values of $\theta_1$ and $\theta_2$, and the ground state manifold has
a Chern number equal to $\pm1$, indicating a topological FQH
phase. The Chern number is a topological invariant, defined as the
integral of the Berry curvature over the $(\theta_1,\theta_2)$ plane.
In practice, we calculate the Berry curvature by dividing the
$(\theta_1,\theta_2)$ plane into a discrete lattice of $N_\theta
\times N_\theta$ points. At each point, the Berry curvature is given
by the Berry phase acquired by the state around a small loop. Then, we
numerically integrate the Berry curvature to obtain the Chern
number. For a more detailed description, see
Ref. \onlinecite{Resta_2011}.

The model systems used in the present work are the checkerboard lattices
with $4\times3$ and $5\times3$ unit cells at filling $\nu=\frac{1}{3}$, i.e. $N_p = 4$
particles on $N_s = 4\times3\times2 = 24$ lattice sites and $N_p = 5$ particles on $N_s = 5\times3\times2 = 30$ lattice sites, respectively. We perturb the system by varying up to five of the
nearest-neighbor hopping amplitudes, i.e. setting $t_{jk} \rightarrow
s$ for some of the hopping amplitudes and $t_{jk} \rightarrow t$ to
all others in the first term of the Hamiltonian (\ref{Ham}), or by
inserting up to five local potentials by adding terms of the form
\begin{equation}
H_j^{pot} = pn_j
\end{equation}
to the Hamiltonian. Negative and positive values of $p$ correspond to
attractive and repulsive potentials, respectively. The locations of
these impurities have been chosen such that they are evenly
distributed in the lattice, as presented in Figure \ref{fig1} for the $4\times3$ unit cell lattice.

To gain insight into the phase transitions caused by the impurities,
we compute the spectrum of the one-body reduced density matrix
(1-RDM)\cite{Coleman_1963} with elements

\begin{equation}
\rho_{ij} = \left<\Psi_0\right|c_i^{\dagger}c_j\left|\Psi_0\right>,
\end{equation}
where $\left|\Psi_0\right>$ is the ground state. The 1-RDM eigenstates
are called natural orbitals (NO). For trivial uncorrelated single
Slater determinant states the eigenvalues of the 1-RDM have values 0
and 1, and for correlated states they fall somewhere in-between,
indicating that the state cannot be fully described by single particle
physics. The sum of the 1-RDM eigenvalues equals the particle number,
and in a FQH state the nonzero eigenvalues are equal to the filling
fraction. Thus, by computing the 1-RDM spectrum we obtain information
about the quantity of correlations present, and furthermore, how the
impurities bind particles from the correlated many-body
state.

We solve the lowest eigenstates and energies by the exact diagonalization method
in the full Hilbert space. We use the Lanczos algorithm to obtain the lowest eigenstates and eigenenergies of the Hamiltonian. In the Lanczos iteration, typically only the lowest eigenstate will be accurate enough, so to obtain the excited states, we run the algorithm multiple times, and shift up the eigenvalues of the states that have already been computed. The shift is done by adding
\begin{equation}
H_m^{\text{shift}} = \lambda \sum_{i=0}^{m-1}\left|\Psi_i \right>\left<\Psi_i \right|
\end{equation}
to the Hamiltonian when computing the $m$th eigenstate $\left|\Psi_m \right>$. This will increase the eigenvalues of the previously computed states by $\lambda$, so as long as $\lambda$ is a large enough positive number, the next Lanczos iteration will converge to the $m$th eigenstate. Our implementation of the Lanczos algorithm runs fully
on a graphics processor, programmed with the CUDA programming
model\cite{Siro_2012}.

\section{Impurities}

\begin{figure*}[ht.]
 \includegraphics[width=0.45\textwidth]{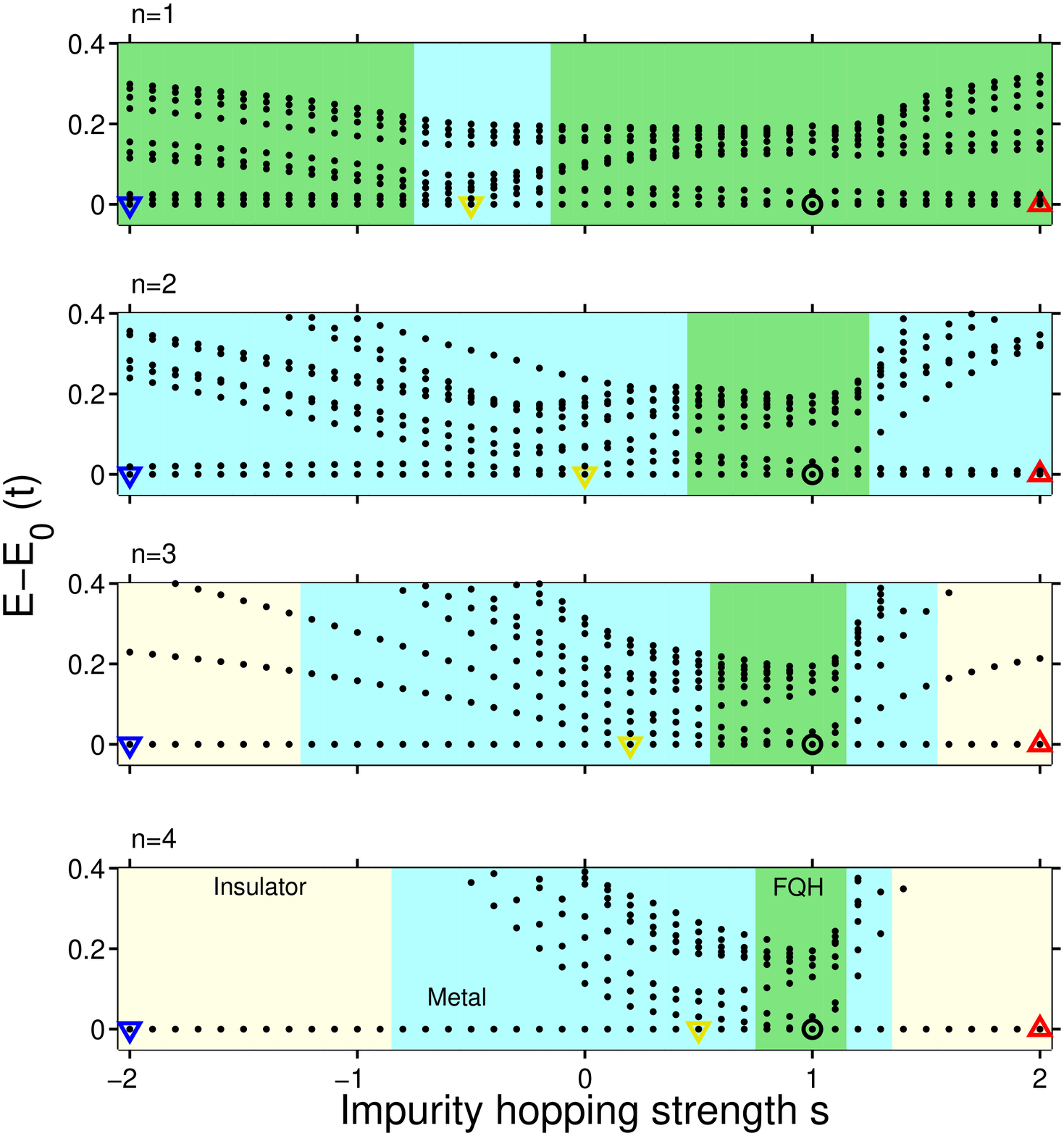}
 \includegraphics[width=0.45\textwidth]{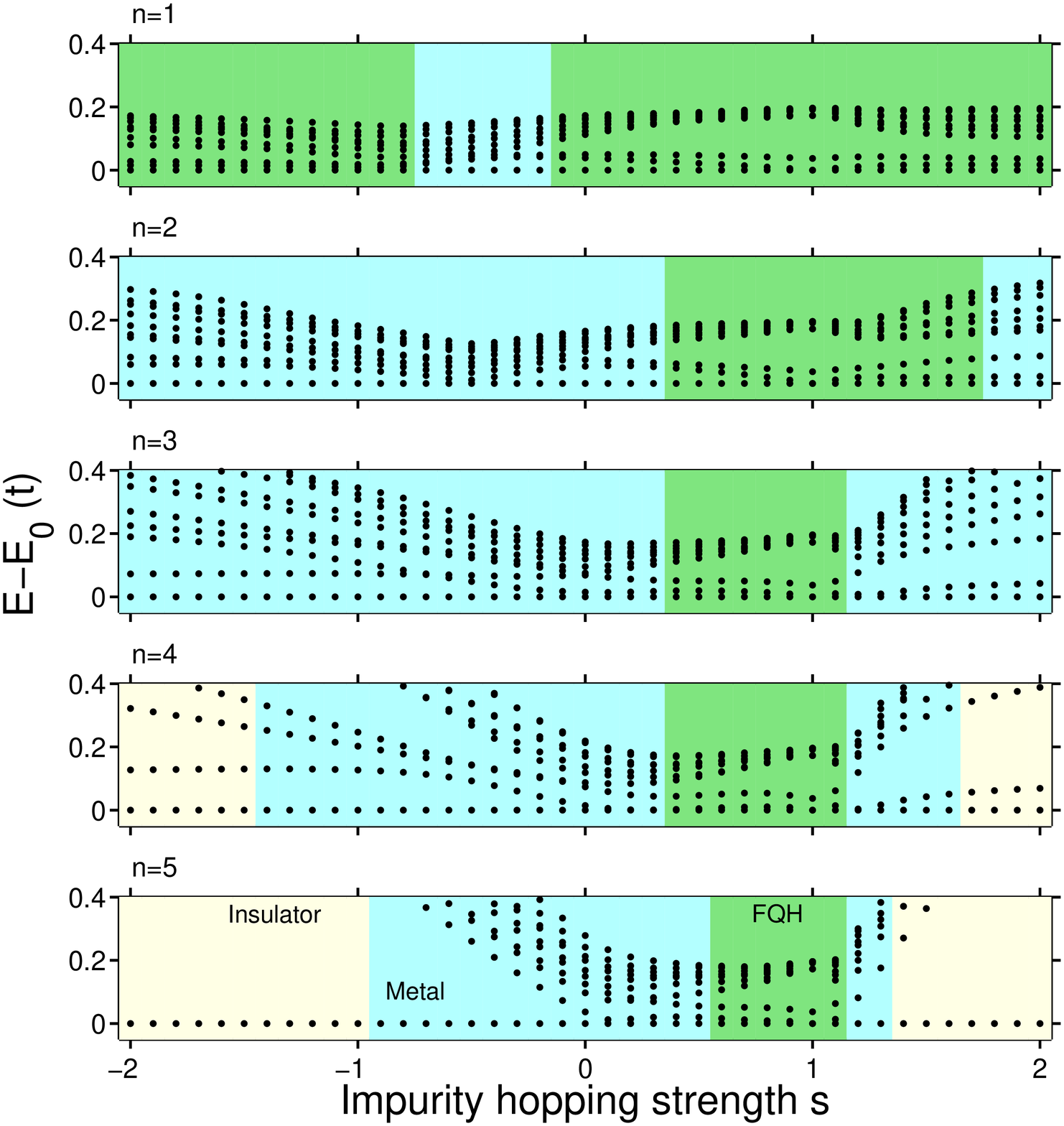}
  \caption{\label{fig4} \raggedright(Color online) Ten lowest
    eigenenergies of the Hamiltonian in Eq. (\ref{Hint}), averaged over the twisted boundary conditions. There are $n=1,2,3,4, 5$ added impurity hoppings at $V=3t$ on $4\times3$ (left) and $5\times3$ (right) unit cell checkerboard lattices. The FQH, metallic and
    insulating phases are indicated by the different background
    colors, see the bottom panels.}
 \end{figure*}

\begin{figure}[ht.]
  \includegraphics[width=.45\textwidth]{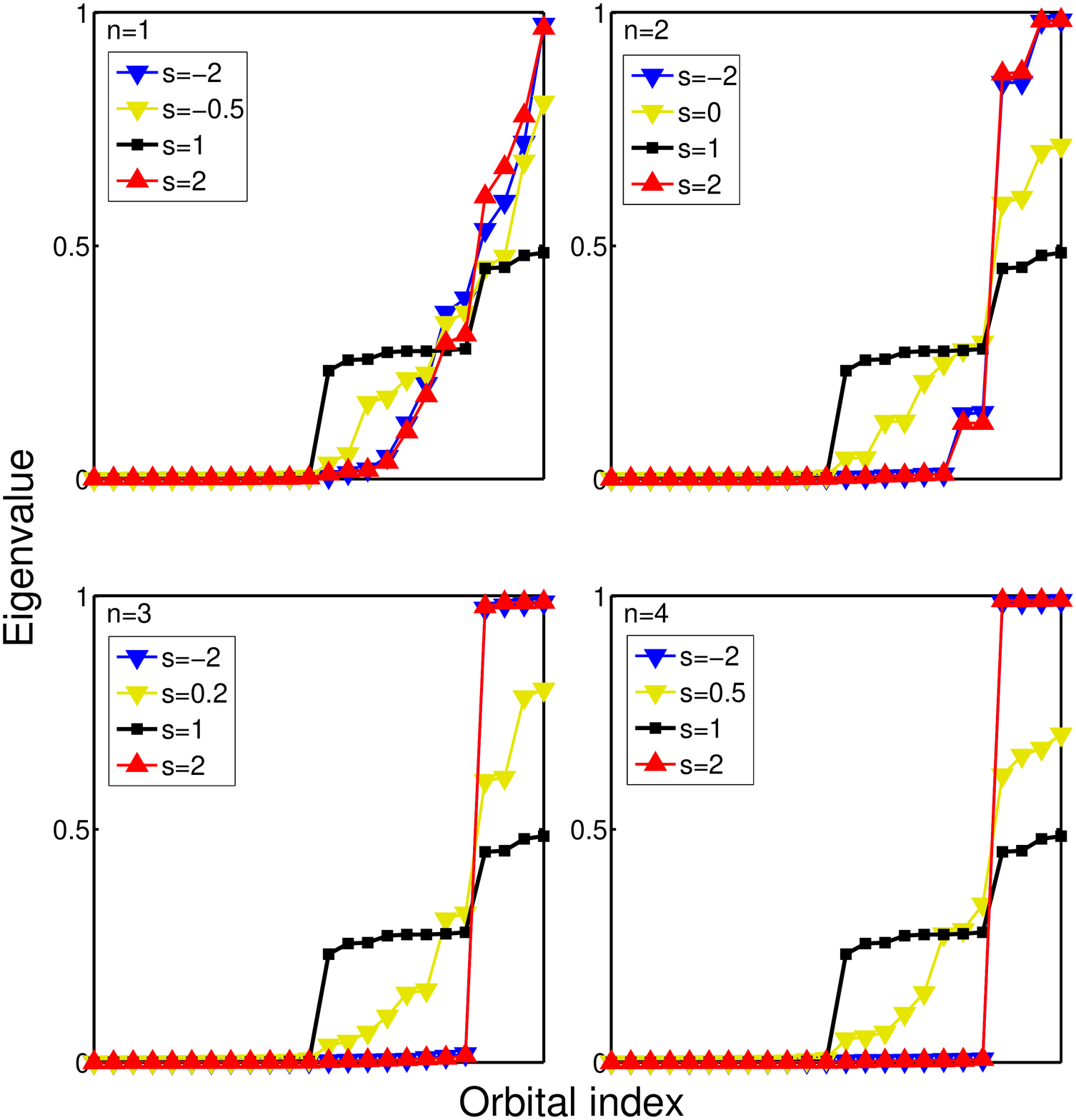}
 \caption{\label{rdmrandt} (Color online)  The spectra of
    the one-body reduced density matrices of the ground states at the
    highlighted points on the left hand side plots of Figure \ref{fig4}.}
 \end{figure}
 
 \begin{figure}[ht.]
 \includegraphics[width=0.4\textwidth]{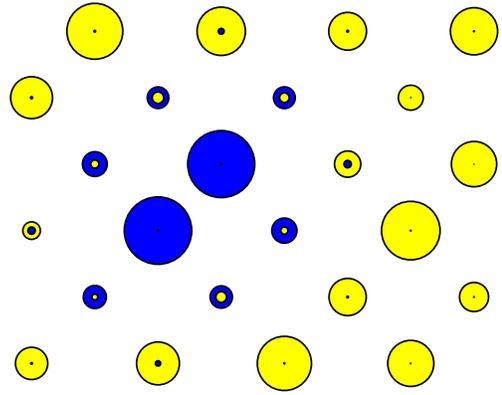}
 \caption{\label{figlocalt} (Color online) Natural orbital densities in the $4\times3$ unit cell checkerboard lattice with $n=1$ impurity hopping
   at $s=-2$. The densities are indicated by the areas of the circles. The highest occupied natural orbital is plotted in blue,
   and the sum of the three next highest occupied natural orbitals is
   plotted in yellow. }
 \end{figure}

%\begin{figure}[ht.]
% \includegraphics[width=0.5\textwidth]{paper_gsm_3randt.eps}
%  \caption{\label{figgsm} The ground state manifold with $n=3$
%    impurity hoppings at $s=1,1.05,1.5$ in the $4\times3$ unit cell checkerboard lattice.}
% \end{figure}

Previously, the effect of adding attractive local potentials to the
checkerboard lattice has been discussed in Ref.
\onlinecite{Yang_2012}. Phase transitions to a metallic state and a
topologically trivial insulating phase were observed to take place
with increasing strength of the potentials. We would like to conduct a
more thorough study of the robustness of the FQHE state against
different kinds of impurities, including a variable concentration of local
potentials and variations in the hopping amplitudes between sites.

We detect the phase transitions caused by the impurities by computing
the low energy and the 1-RDM spectra. The FQH phase is clearly
identified by the threefold quasi-degenerate ground state manifold
with a unit total Chern number, separated by a large gap from the
higher energy states. Transitions from the topological FQH phase into
a metallic phase are characterized by the collapse of the energy gap above
the GSM. The insulating phase has one ground state that is separated
from other states by a large gap in the whole $(\theta_1,\theta_2)$
plane. It is also topologically trivial, indicated by the vanishing
Chern number of the ground state.

The 1-RDM spectra can be used to corroborate the analysis: the
localization of the particles onto the impurities leads to fully
occupied natural orbitals, i.e. unit eigenvalues. In the highly
correlated FQH phase, we observe non-integer eigenvalues whereas in the
trivial insulator phase all nonzero eigenvalues are very close to
unity.

Explicitly, we use the following criteria to determine the phase: the system is in the FQH phase if the GSM has a unit total Chern number and the energy gap above the GSM is larger than the width of the GSM, averaged over the twisted boundary conditions. If the ground state has a vanishing Chern number and its 1-RDM spectrum has as many eigenvalues larger than $0.95$ as the particle number, the system is in the insulating phase. Otherwise, the system is in the metallic phase. These criteria are of course somewhat arbitrary, but fixing them allows us to compare the phases in systems of different sizes. 

\subsection{Impurity hoppings}
We will now present our results for the $4\times3$ and $5\times3$ unit cell checkerboard lattices with added impurity hoppings.
The energies of the ten lowest states at nearest-neighbor interaction strength $V=3t$ 
are presented in Figure \ref{fig4}. They have been averaged over the twisted boundary conditions.
The plots show the energies as a function of 
the impurity hopping strength $s$, and $n=1,2,3,4,5$ indicates the number of
impurity hoppings. The energy spectra on the left are from the $4\times3$ system with up to four impurity hoppings and on the right from the $5\times3$ system with up to five impurity hoppings. Additionally, in Figure \ref{rdmrandt}, we present the 1-RDM spectra for the ground state of the $4\times3$ system at the points highlighted by the colored markers in Figure \ref{fig4}. The 1-RDM spectra for the $5\times3$ system are essentially identical.
% In addition to the unperturbed
%state, each plot has the 1-RDM spectra at three other values of
%impurity strength, corresponding to the highlighted points in the
%energy spectrum plots.
 
With one modified hopping, i.e. $n=1$, we observe an interesting
energy spectrum for both system sizes. The ground state manifold and the gap to higher
energy states seem to stay intact for $s>1$. However, with $s<1$ the
gap closes with a small negative $s$ but reopens with a larger negative $s$. For large
negative values of $s$, there is a new threefold quasi-degenerate
ground state manifold. The total Chern number of this new GSM is unity,
so it seems that the FQH phase re-emerges at large negative $s$. The
peculiar metallic phase in between has a total Chern
number two so it could also be interpreted as two different FQH phases
reorganizing as the impurity hopping changes sign.

The $n=1$ 1-RDM spectrum in Figure \ref{rdmrandt}
further supports the re-emergence of the FQH phase with a different
GSM as seen in the corresponding energy plot. First of all, the 1-RDM spectrum of the
unperturbed system at $s=1$ is plotted in black. The first 12
eigenvalues are essentially zero, and they correspond to the states in
the higher band in our non-interacting two-band model. The rest of the
eigenvalues have non-integer values, as expected from the highly
correlated $\nu=\frac{1}{3}$ FQH state.
 The 1-RDM spectra of the
states with strong impurities, $s=2$ and $s=-2$, are very similar. The shape reflects the
division of the lattice into two distinct areas: because of the large
hopping amplitude, a single electron is localized to the sites that
are connected by the impurity, seen in the plot as the one almost
fully occupied orbital. This localization, in turn, causes the
occupation of the neighboring sites to decrease to avoid the cost of
the repulsive NN interaction, indicated by the decreased occupancy in
some NOs. In the remaining lattice, the rest of the particles form an
FQH type phase, characterized by the non-integer part of the spectrum.

This interpretation is confirmed by looking at the natural orbital
densities for each lattice site. In Figure \ref{figlocalt}, the fully
occupied natural orbital is mostly localized on the two lattice sites
connected by the impurity hopping, with some density on the
neighboring sites and essentially zero density elsewhere. On the other
hand, the three next highest occupied NOs form a somewhat even density
in the remaining lattice, with very little density at the impurity
site or its nearest neighbors.

With two impurity hoppings, transitions to the metallic phase occur
for both negative and positive $s$. The 1-RDM spectra show two fully
occupied NOs which are localized on the the two impurity
hoppings. Again, the spectra for $s=2$ and $s=-2$ are almost identical
as the sign of the impurity hopping only affects the localized NOs,
since the density of the other NOs vanishes on the sites connected by
the impurities.

As the impurity density increases, the FQH region around the unperturbed
system at $s=1$ becomes narrower. With $n\geq3$ for the $4\times3$ system and $n\geq4$ for the $5\times3$ system, strong impurities cause a transition
into the trivial insulating phase with one ground state separated from
the others by a very large gap. This is because most or all particles
have been localized on the impurities. The corresponding 1-RDM
eigenvalues for $s=2$ and $s=-2$ are all very close to zeros and ones,
confirming that the ground state is almost fully uncorrelated.

\begin{figure*}[ht.]
 \includegraphics[width=0.45\textwidth]{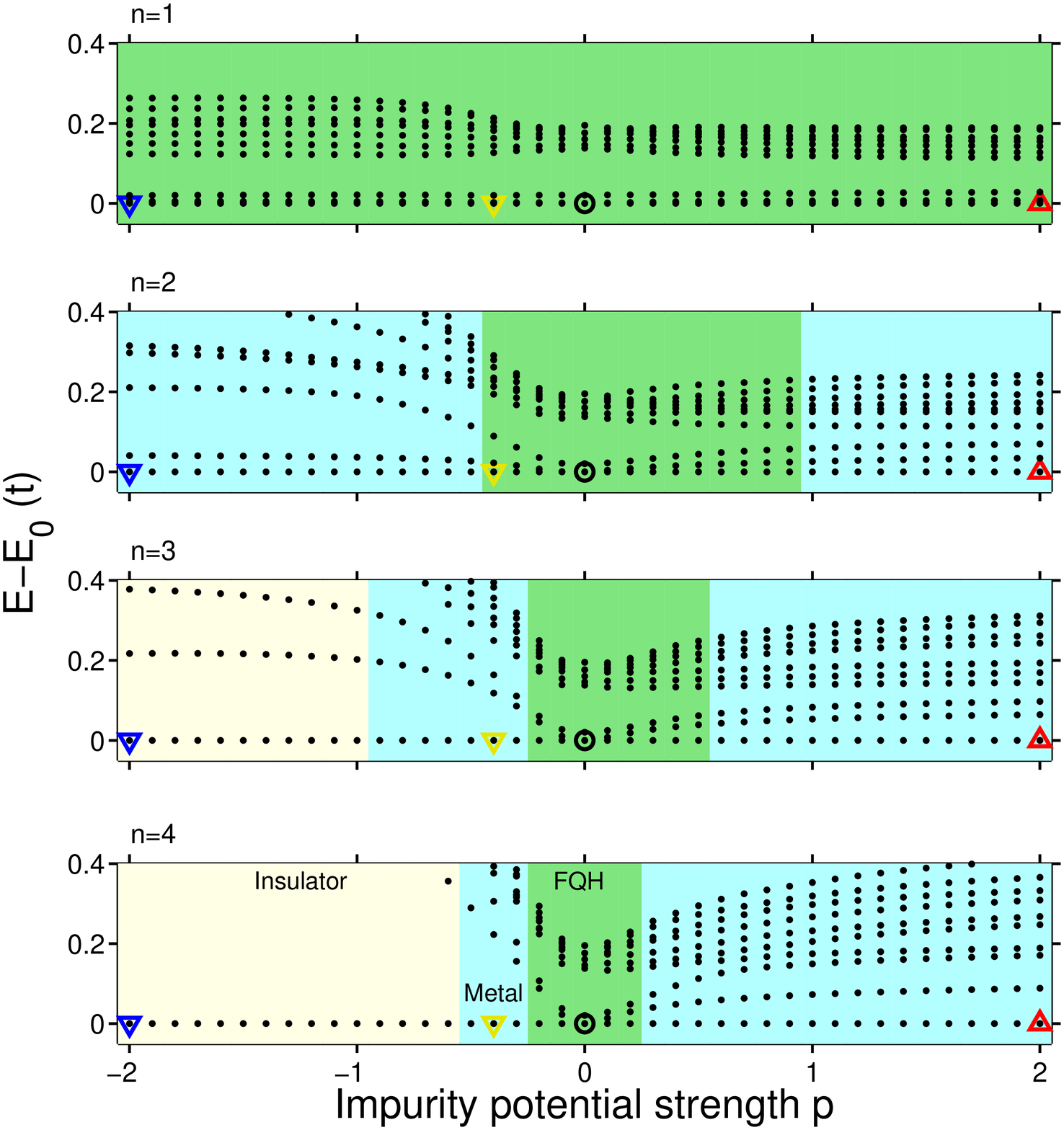}
 \includegraphics[width=.45\textwidth]{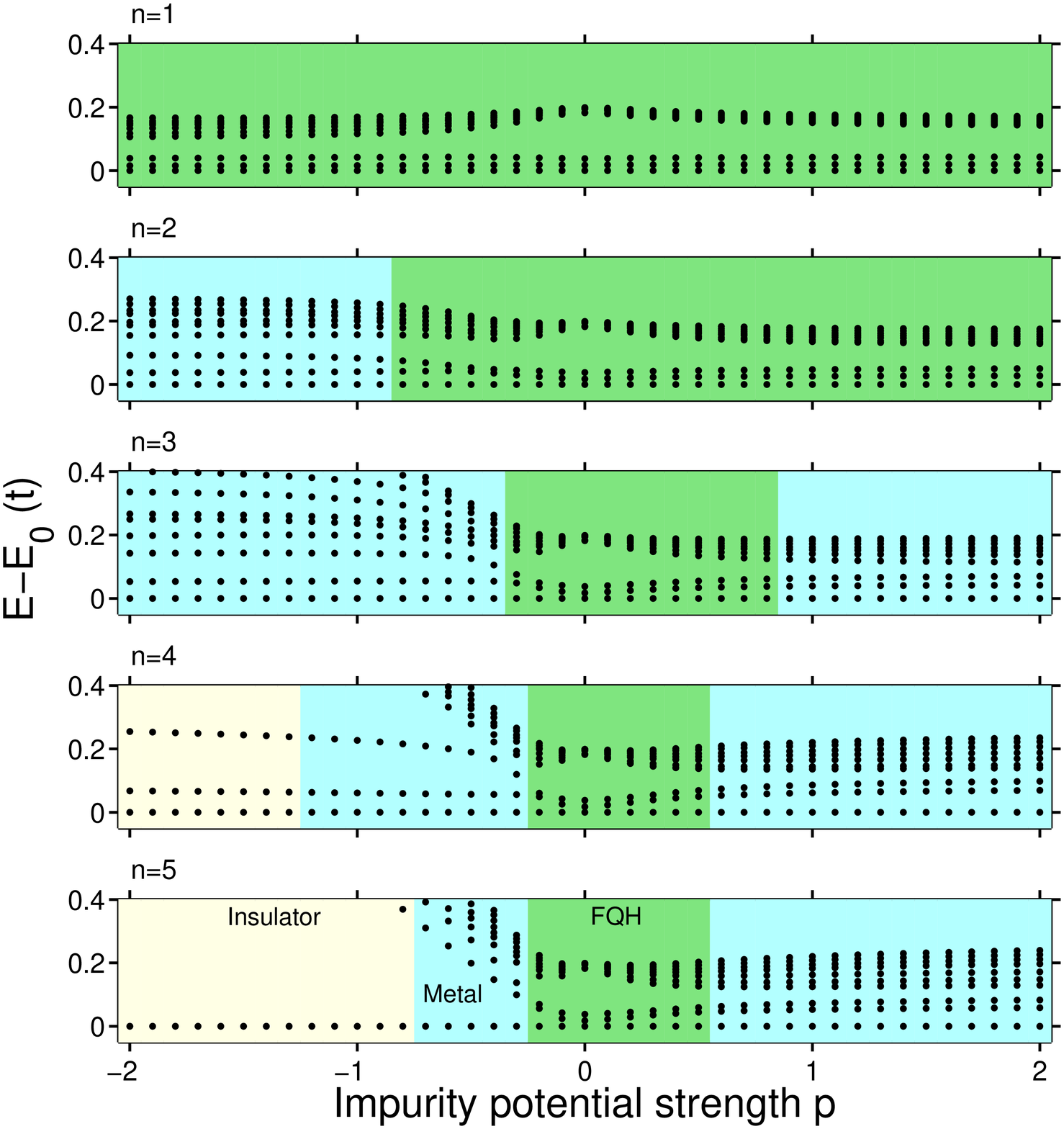}
  \caption{\label{fig5}\raggedright (Color online) Ten lowest
    eigenenergies of the Hamiltonian in Eq. (\ref{Hint}), averaged over the twisted boundary conditions. There are
    $n=1,2,3,4,5$ added impurity potentials at $V=3t$ on $4\times3$ (left) and $5\times3$ (right) unit cell checkerboard lattices. The FQH, metallic and
    insulating phases are indicated by the different background
    colors, see the bottom panels.}
 \end{figure*}
 
 \begin{figure}[ht.]
  \includegraphics[width=.45\textwidth]{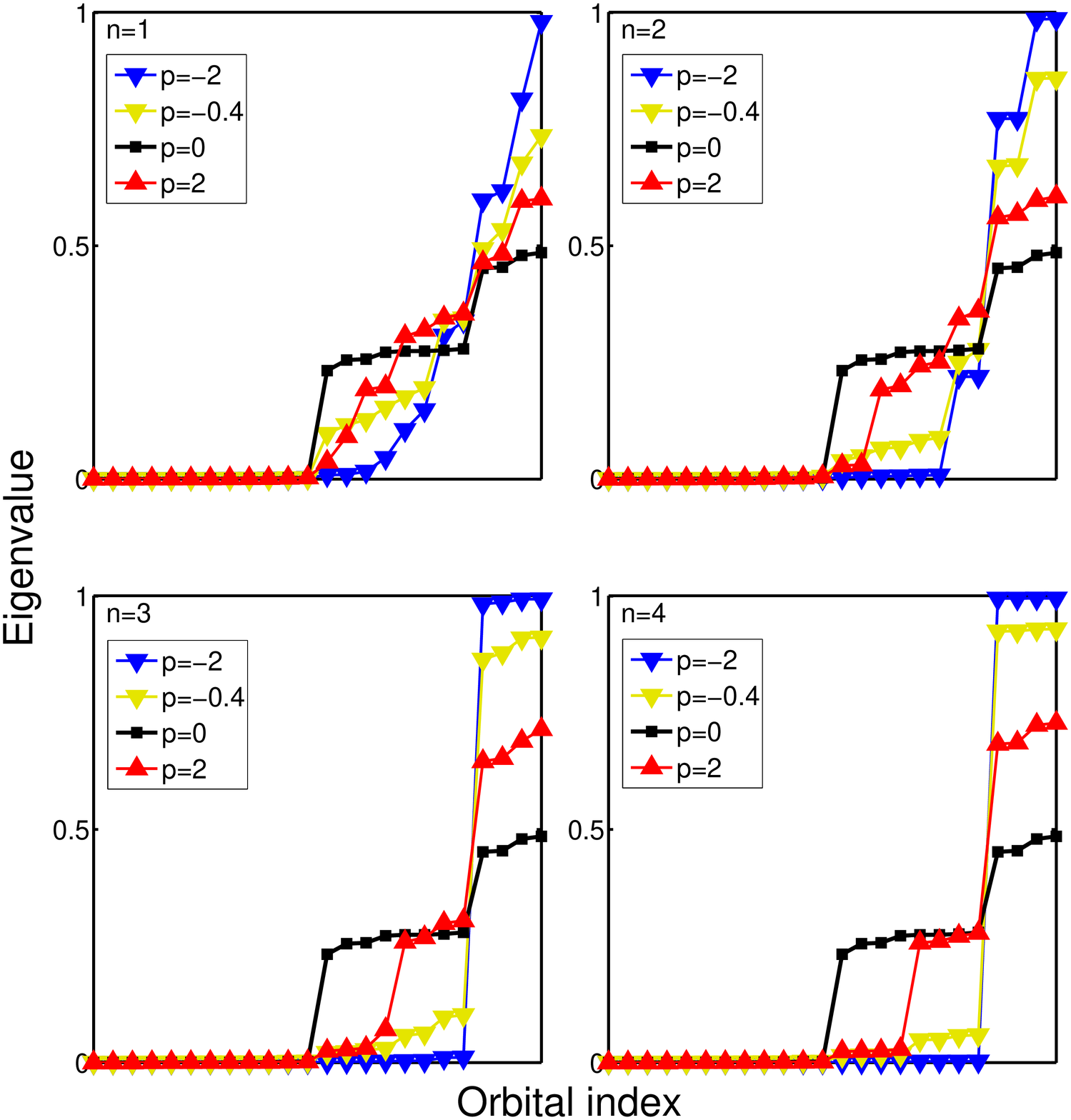}
 \caption{\label{rdmpot} (Color online) The spectra of
    the one-body reduced density matrices of the ground states at the
    highlighted points on the left hand side plots of Figure \ref{fig5}. }
 \end{figure}

\begin{figure}[ht.]
 \includegraphics[width=0.4\textwidth]{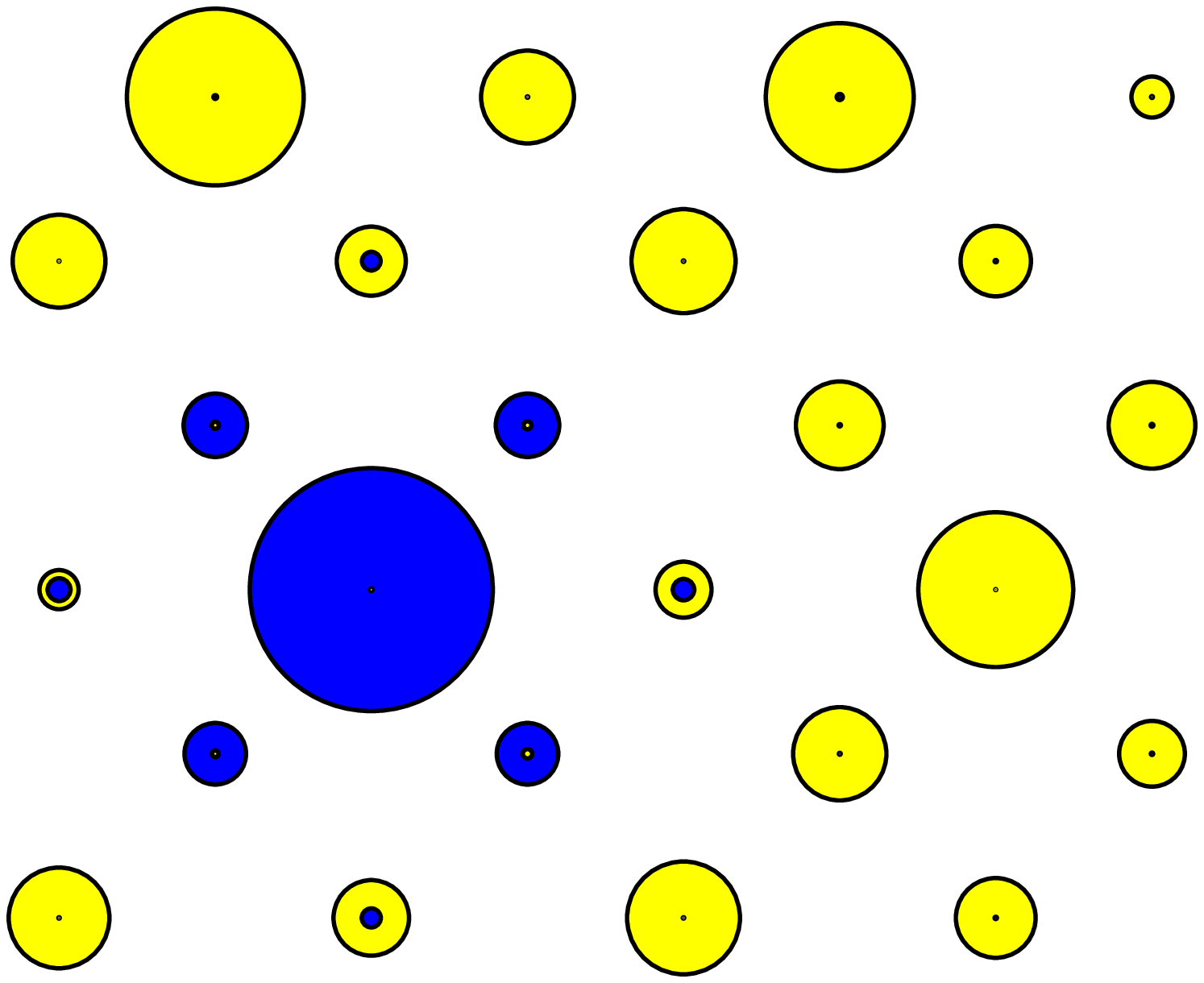}
 \caption{\label{figlocalp} (Color online) Natural orbital densities in the $4\times3$ unit cell checkerboard lattice with $n=1$ impurity
   potential at $p=-2$.  The densities are indicated by the areas of the circles. The highest occupied orbital is plotted in
   blue, and the sum of the three next highest occupied natural
   orbitals is plotted in yellow. }
 \end{figure}

%\begin{figure*}[h!.]
% \includegraphics[width=0.45\textwidth]{paper_enes_12345randt.eps}
% \includegraphics[width=.45\textwidth]{paper_enes_12345randpot.eps}
%  \caption{\label{figenes5x3}\raggedright (Color online) Ten lowest
%    eigenenergies of the Hamiltonian in Eq. (\ref{Hint}) with
%    $n=1,2,3,4, 5$ added impurity hoppings (left) and impurity potentials (right) at $V=3t$ on a $5\times3$ unit cell checkerboard lattice. The FQH, metallic and
%    insulating phases are indicated by the different background
%    colors, see the $n=5$ panel.}
%    \vspace{0.5cm}
% %\end{figure*}
% 
% %\begin{figure*}[ht.]
%% \includegraphics[width=0.9\textwidth]{paper_rdms_12345randt.eps}
%% \includegraphics[width=0.9\textwidth]{paper_rdms_12345randpot.eps}
%%  \caption{\label{figrdms5x3}\raggedright (Color online) The spectra of
%%    the one-body reduced density matrices of the ground states at the
%%    highlighted points in Figure \ref{figenes5x3}. The top row shows the spectra for the impurity hoppings and the bottom row for the impurity potentials. }
% \end{figure*}

\subsection{Impurity potentials}
Analogous results for the systems with up to five added impurity 
potentials are presented in Figures \ref{fig5}, \ref{rdmpot} and \ref{figlocalp}.  With one potential, we
observe no phase transitions: the ground state manifold and the gap
remain intact with both repulsive and attractive perturbations. Unlike
with the single impurity hopping, there are no changes in the states
in the GSM. The $n=1$ 1-RDM spectra show that at $p=-2$ there is a
fully occupied natural orbital, which is localized on the lattice site
with the attractive potential.

As above with the impurity hoppings, the localization can be directly
seen in the densities obtained from the most occupied NOs, presented
in Figure \ref{figlocalp}. The fully occupied NO has most of its
density in the lattice site with the attractive potential, while the
three next highest occupied NOs are distributed elsewhere in the
lattice, with a very small density near the impurity potential. Again,
the conclusion is that the system is divided into the localized part
near the impurity site and the FQH state elsewhere. Interestingly,
there is a very smooth transition from the FQH state formed by four
particles at $p=0$ to the one formed by three particles at $p=-2$.

In the $4\times3$ system with two potentials, there are phase transitions to the metallic state
with both attractive and repulsive impurities. Unlike in the case with
impurity hoppings, there is a clear asymmetry between positive and
negative values of $p$, clearly seen in the 1-RDM spectra. The strong
attractive potentials at $p=-2$ have bound two particles to the
impurities, seen in the two fully occupied NOs. On the other hand, the
1-RDM spectrum for the ground state with strong repulsive potentials
at $p=2$ is quite close to the unperturbed FQH ground state at
$p=0$. This is reasonable, since from the energy spectra one can see
that the effect of the repulsive potentials is quite small: they only
widen the GSM and make the gap smaller. In the $5\times3$ system, two repulsive potentials are not enough to break the FQH phase. This is most likely because the impurity concentration is smaller than in the $4\times3$ lattice, which results in the gap being more robust.

In the smaller system with three and four potentials, there is a quick phase transition into the
metallic phase and onto the insulating phase with attractive
potentials, while repulsive potentials only induce a transition to the
metallic phase. Again, in the larger system the FQH phase seems to be more resilient against repulsive potentials, as the gap above the GSM stays open with larger positive $p$ than in the $4\times3$ system.  However, with attractive potentials, we observe the same transition to the insulating phase through the metallic phase. 

%The 1-RDM spectra confirm that in our small system,
%three attractive potentials are enough to remove correlations from the
%system, as all four highest occupied NOs have eigenvalue very close to
%unity.

The cases with equal numbers of attractive impurity potentials and particles, i.e. $n=4$ in the $4\times3$ system and $n=5$ in the $5\times3$ system with negative $p$, correspond to the one studied in
Ref. \onlinecite{Yang_2012}. Qualitatively, our results agree: there
are two phase transitions, one from the FQH to the metallic phase, and
another from the metallic phase to the insulating phase. We can
clearly identify these phases in the corresponding 1-RDM
spectra. There are quantitative distinctions, probably due to
differences in system size and the location of the impurity
potentials.

\subsection{Finite size scaling}
In the previous sections, we have presented results for two system sizes with $4 \times 3$ and $5 \times 3$ unit cells, respectively. The results are in qualitative agreement: the FQH state is robust against individual impurities and the FQH region becomes narrower with increasing concentration of impurities. The insulating phase is only obtained for strong impurity hoppings or strong attractive potentials when the impurity concentration is close to the filling fraction. To perform a rudimentary finite size scaling, we have also done computations on a $6 \times 3$ unit cell lattice with $N_p = 6$ particles. 

\begin{figure}[t]
 \includegraphics[width=0.21\textwidth]{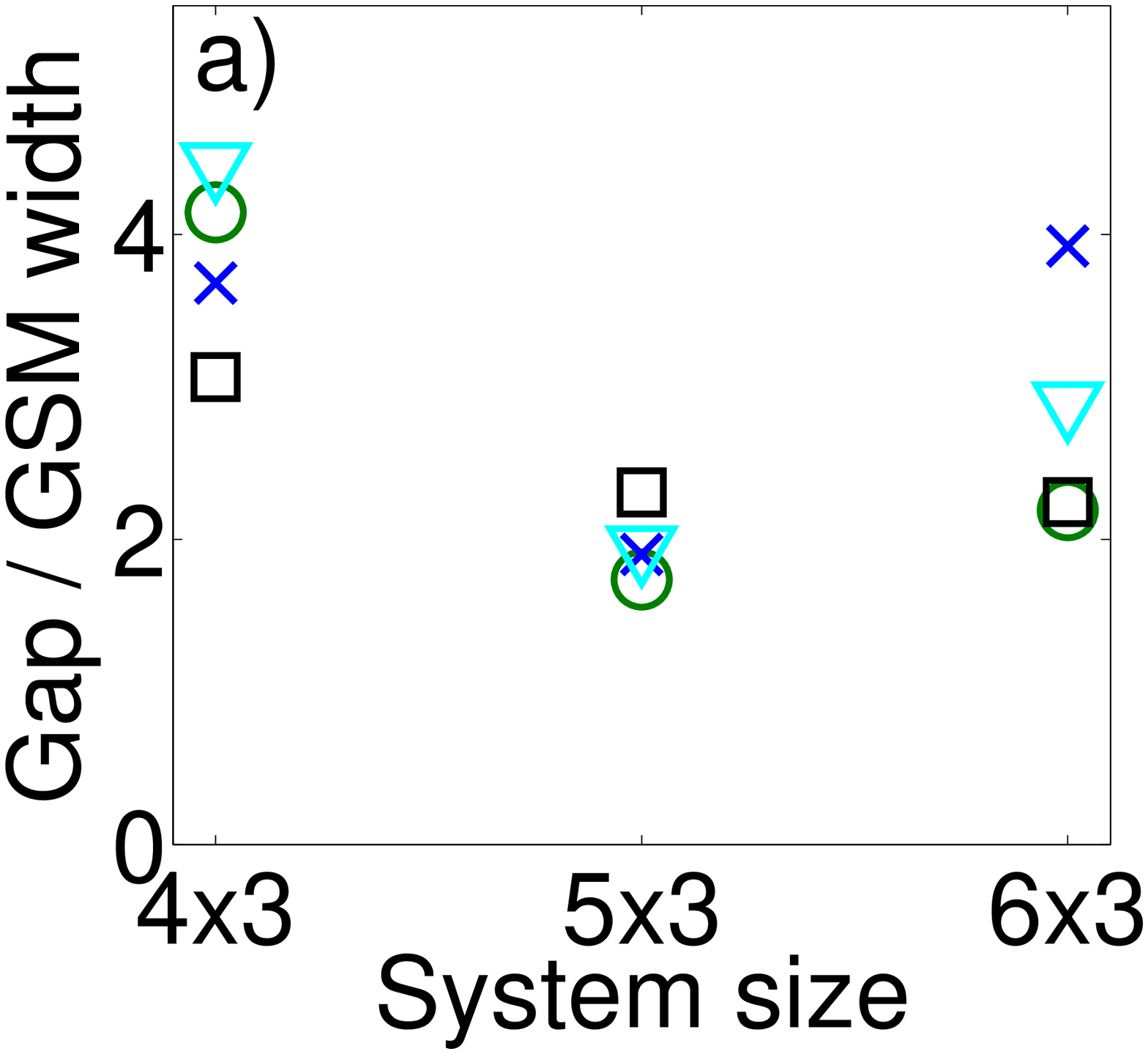}
  \includegraphics[width=0.21\textwidth]{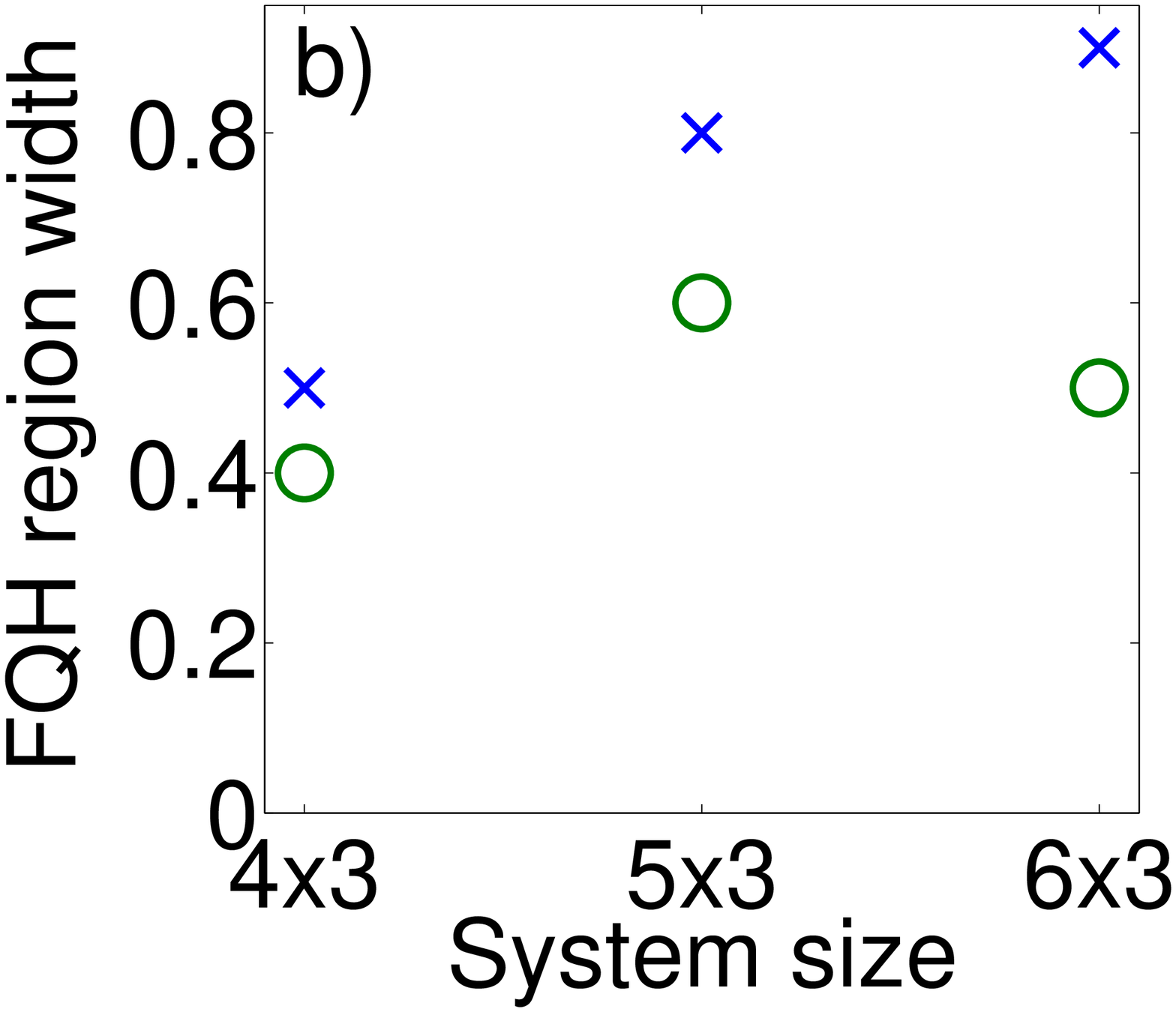}

  \caption{\label{figscaling}(Color online) (a) The ratio of the energy gap above the GSM to the width of the GSM as a function of the system size in four cases: one impurity hopping at $s=-2$ (blue crosses) and $s=2$ (teal triangles) and one impurity potential at $p=-2$ (green circles) and $p=2$ (black squares). (b) The width of the FQH phase region with impurity hoppings (green circles) and impurity potentials (blue crosses) as a function of the system size. The number of the impurities is equal to the number of particles.  }
 \end{figure}

The results in Figure \ref{figscaling} demonstrate the robustness of the FQH phase as a function of the lattice size for both low and high number of impurities. In Figure \ref{figscaling}a, we plot the energy gap above the GSM divided by the width of the GSM with a single impurity hopping at strength $s=\pm2$ and a single attractive potential with strength $p=\pm2$. Note that here the impurity density decreases as the system becomes larger. At the thermodynamic limit, one would expect the FQH phase in the case of an individual local impurity. Unfortunately, we are only able to increase one of the dimensions in the system size because of the size limitation of the exact diagonalization method and the fact that the number of unit cells needs to be divisible by three to be able to support a $\nu=\frac{1}{3}$ FQH phase. Thus, properly extrapolating to the 2D thermodynamic limit is beyond the scope of this study. In Figure \ref{figscaling}b, we plot the width of the FQH phase region in the presence of impurity hoppings and potentials, such that the number of impurities is equal to the number of particles. Here, the impurity density stays constant. Despite the thermodynamic limit being out of reach, it seems plausible that the remarkable robustness of the FQH phase extends also to larger systems.

%All the results presented in the previous sections have been computed for the $4 \times 3$ unit cell checkerboard lattice. To study the finite size effect and confirm that the $4\times3$ unit cell lattice is not just a special case, we have also computed the energy and 1-RDM spectra for a $5 \times 3$ unit cell lattice with $5$ particles and up to $5$ impurity hoppings and potentials, see Figures \ref{figenes5x3} and \ref{figrdms5x3}. The results are in qualitative agreement: the FQH state is robust against individual impurities and the FQH region becomes narrower with increasing concentration of impurities. The insulating phase is only obtained for strong impurity hoppings or strong attractive potentials when the impurity concentration is close to the filling fraction. In the larger system, the FQH phase seems to be even more robust against repulsive potentials, indicated by the clear gap above the GSM in the $n=2$ and $n=3$ plots on the right hand side of Figure \ref{figenes5x3}. 

\section{Landau level mixing}

\begin{figure}[t]
 \includegraphics[width=0.45\textwidth]{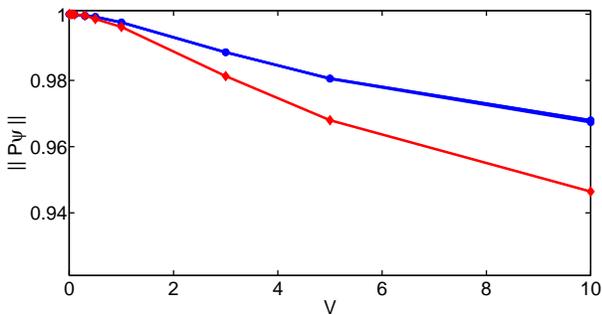}
  \caption{\label{fig2}(Color online) The norm of the three ground
    states (in blue) and the first excited state (in red) of the $4 \times 3$ unit cell checkerboard lattice after
    projecting them onto the LLL as a function of the nearest-neighbor
    interaction $V$, in units of $t$.  }
 \end{figure}
 
 \begin{figure}[t]
 \includegraphics[width=0.45\textwidth]{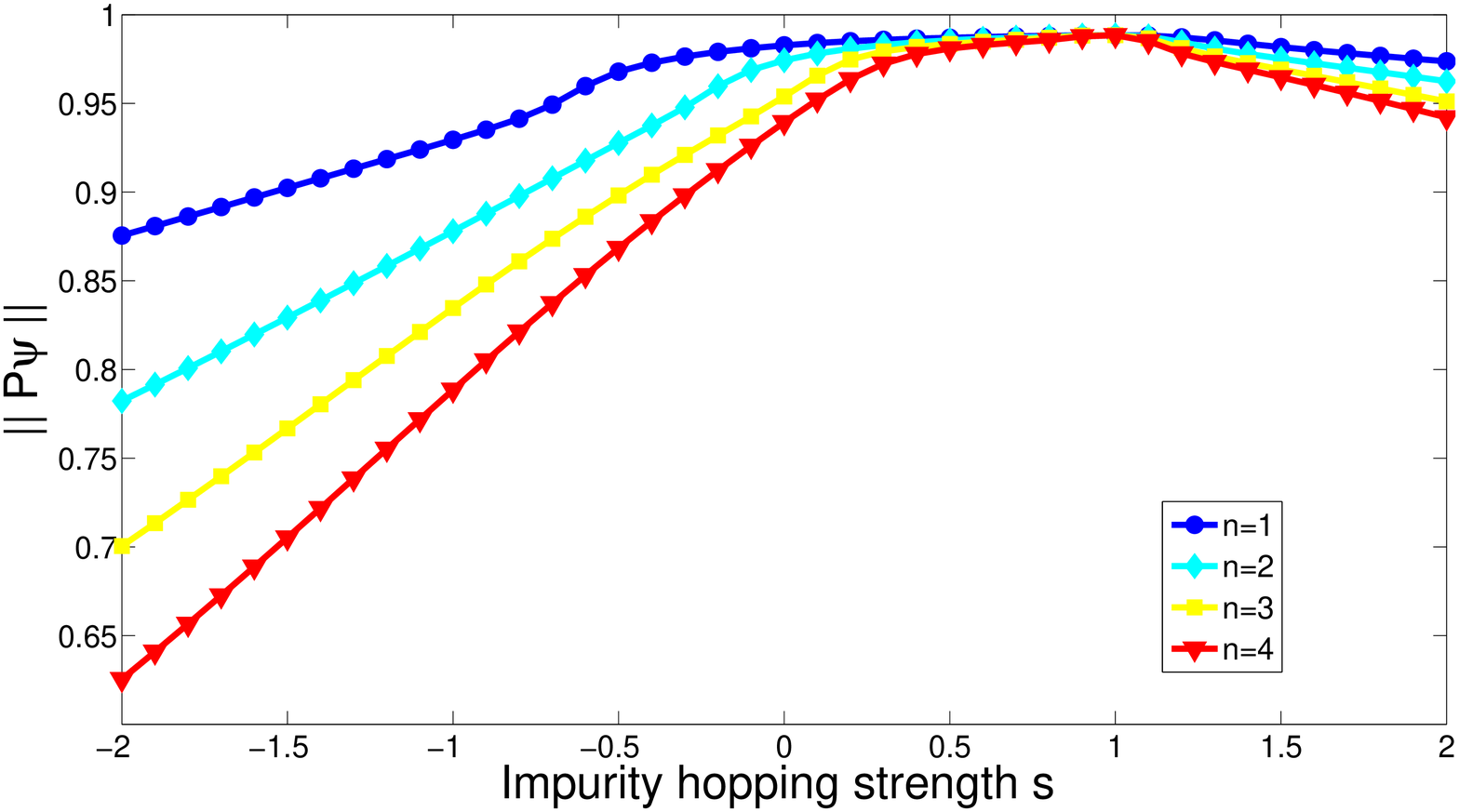}
  \caption{\label{fig6}(Color online) The norm of the ground states of the $4 \times 3$ unit cell checkerboard lattice
    with $n$ impurity hoppings after projecting them onto the LLL as a
    function of the impurity strength.  }
 \end{figure}
 
 \begin{figure}[t]
 \includegraphics[width=0.45\textwidth]{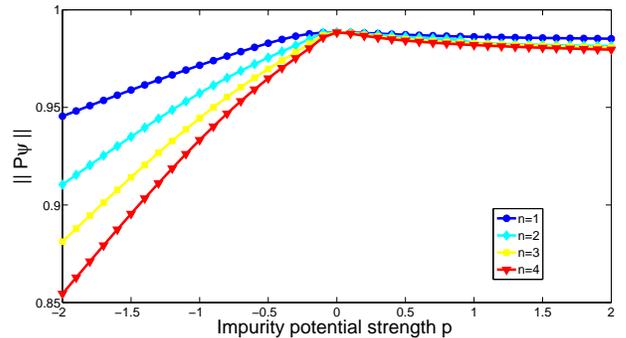}
  \caption{\label{fig7}(Color online) The norm of the ground states of the  $4 \times 3$ unit cell checkerboard lattice
    with $n$ impurity potentials after projecting them onto the LLL as
    a function of the impurity strength.  }
 \end{figure}

A feature in the 1-RDM spectra is the fact that in
all cases the twelve lowest eigenvalues are nearly zero, indicating
vanishing occupancy of half of the natural orbitals. In the two-band
checkerboard model, the flat energy band corresponds to the lowest
Landau level (LLL) in the ordinary continuum quantum Hall effects. An
explanation to these unoccupied natural orbitals is that they consist
mostly of the excited single particle states lying above the nearly
degenerate flatband states. This observation seems to imply very
little mixing to the higher energy states even with strong impurities.

For example in Ref. \onlinecite{Parameswaran_2013}, it has been
conjectured that to obtain the FQH state, one should have the
following hierarchy of energy scales:
\begin{equation}
\label{hierarchy}
b \ll V \ll \delta,
\end{equation}
where $b$ is the bandwidth of the flat band, $V$ is the interaction
and $\delta$ is the band gap.  In the literature, computations have
often been limited to the low-energy band by making it completely flat
and projecting the states onto it, essentially making the band gap
infinite\cite{Wu_2012,Bernevig_2012}. This allows for a smaller
Hilbert space which makes computations easier and also ensures that
the above criterion is fulfilled. However, it seems that it is not
necessary to obey the hierarchy of Eq. (\ref{hierarchy}) as our
results were computed at $V=3t$, which is certainly larger than the
band gap in the checkerboard model. Moreover, in
Refs. \onlinecite{Kourtis_2013,Kourtis_2012}, much higher values of interaction have
been used successfully.

Since we have performed our calculations without any approximations to
the Hilbert space, we can study how much mixing to higher states is
caused by the interaction $V$ and the various impurities. This is
interesting from the point of view of a potential experimental
realization of the model. Previously, mixing of bands with nonzero Chern numbers has been studied in a 
triangular lattice in Ref. \onlinecite{Kourtis_2013_2}.

We measure the extent of the mixing by projecting the computed ground
states onto the lowest band with the projection operator
\begin{equation}
P = \sum_j\left|\phi_j\right>\left<\phi_j\right|,
\end{equation}
where $\left|\phi_j\right>$ are the noninteracting Slater determinant
states, formed from all combinations of the single particle states of
the LLL. We can then measure the degree of mixing to higher states by
computing the norm of the vector $P\left|\psi_0\right>$. A unit norm
indicates that the state is completely contained within the lowest
Landau level.

In Figure \ref{fig2}, we present the norms of the three quasi-degenerate ground states and the first excited state as a function of
the nearest-neighbor interaction $V$. We see that even for much larger
interaction strengths than the gap, which is on the order of $t$,
there is very little mixing to the higher states. There is practically
no difference between the three ground states, while the first excited
state mixes slightly more.

To see whether the impurities cause mixing, we have also performed the
projection onto the lowest band on the ground states of the perturbed
systems. The norms of the resulting vectors are presented in Figure
\ref{fig6} (impurity hoppings) and Figure \ref{fig7} (impurity
potentials).  The system is very robust against a single impurity
hopping or potential, indicated by large norms even for strong
perturbations. With increasing number of impurities, the mixing to the
higher band is increased. However, it seems that there is little
correlation between the norm of the projected vector and the phases
obtained from the energy and 1-RDM spectra. All in all, even with many
strong impurities, the ground states still mostly reside in the space
formed by the states in the lowest Landau level.

\section{Conclusions}

In conclusion, we have performed an exact diagonalization calculation
on the checkerboard lattice model with added
impurities in the form of multiple modified nearest neighbor hoppings
and local potentials. We computed the low energy spectra as well as
the spectra of the one-body reduced density matrices of the ground
states to identify the quantum phase transitions from the
$\nu=\frac{1}{3}$ fractional quantum Hall phase into a metallic phase
and a topologically trivial insulating phase.

The FQH state was found to be very robust against single impurities of
both kind, as expected of a topological state that should be stable
against local perturbations. Increasing the number of impurities caused a transition to the
metallic state. When the number of impurity hoppings was close to the particle number, a
transition to the trivial insulating state was observed with both
positive and negative hopping amplitudes. On the other hand, only
attractive potentials were able to induce the same transition and the FQH phase was overall more robust against repulsive potentials.

Common
to both impurity types, the FQH region became narrower with
increasing impurity density. Qualitatively similar results were obtained for three finite systems of $4\times3$, $5\times3$ and $6\times3$ unit cells with $4$, $5$ and $6$ particles and up to $4$, $5$ and $6$ impurities, respectively. Based on our results, it seems that the
$\nu=\frac{1}{3}$ FQH phase in the checkerboard lattice can withstand
both very strong isolated impurities and a high density of weaker impurities, which is a reassuring result for any
attempts at an experimental realization of the system.

We also showed that very little mixing between the single particle
bands is induced by the nearest neighbor interaction or
impurities. This validates the approximation used in many previous
studies of restricting the Hilbert space to only the states in the
lowest Landau level.

\acknowledgments T.S acknowledges financial support from the Finnish
Doctoral Programme in Computational Sciences FICS. This research has
also been supported by the Academy of Finland through its Centres of
Excellence Program (project no. 251748).  We acknowledge the
computational resources provided by Aalto Science-IT project and
Finland's IT Center for Science (CSC).

\bibliography{Topi.bib}

\end{document}